# Displays and antiferroelectric liquid crystals.

O. Hudak
*Technical University Kosice, Faculty of Aerodynamics, Department of Aviation Technical Studies, Rampova 7, SK – 040 01 Kosice*

M. Hudak
*Stierova 23, SK – 040 23 Kosice*

J. Tothova
*Lab, Stierova 23, SK – 040 23 Kosice*

Recently we have formulated theory of bi-layer-dimerized chiral liquid antiferroelectric crystals [1]. Liquid crystal television display panel has application f.e. in tactical aircraft. Its key advantages are: (1) high contrast in small and large areas, (2) shade capability under all levels of illumination including direct-sunlight, (3) uniform high resolution over the entire display area, (4) interface similar to CRT TV display and (5) low power, weight, volume. Cockpit installations have been designed for the display which permit viewing under day and night conditions, see in [2]. While this display panel application description is older as noted in [3] 34 years following their appearance in wristwatches and calculators, liquid crystal display (LCD) technologies now find commercial application in 100-inch TV screens. Several problems remain to be solved. One of the most serious of these is slow response time, which is inevitable in standard, so-called nematic LCDs. Ferroelectric liquid crystals (FLCs) and antiferroelectric liquid crystals (AFLCs) represent attempts to address this time problem. In last few years [4], the unique features of antiferroelectric liquid crystals (AFLCs) have been explored to develop high-end displays. A number of passive- and active-matrix prototypes have been presented. However, although their use in a number of application areas has been suggested, no commercial products have been announced yet. The work [4] reviews the state of the art of AFLC displays, the reasons for their present low incidence in display markets, and the latest developments aiming to overcome the main shortcomings that hinder their development. The world market for displays amounts roughly 50.000 M USD per year, of which about 50% correspond to flat panel displays (FPDs) [5]. From the electrooptical point of view, AM-TFT LCDs are superior to passive matrix devices, for the LC material response can be made independent of the display electronic driving, and specifically of the multiplexing rate. This means that, unlike previous multiplexed passive displays like supertwist nematics, the device resolution - the product resolution frame rate - can be increased without impairing other display characteristics such as brightness [5]. Antiferroelectric liquid crystals (AFLCs) are a type of liquid crystals ($SmC*A$ ) [6-10] in which spontaneous polarization has been created on different layers due to the helical superstructure. The c-director of the adjacent layers of AFLCs is nearly antiparallel to each other because of its anticlinic nature. It is not perfectly antiparallel as a slow precession of the tilt plane of themolecules fromlayer to layerwhich has been interpreted the chirality in the system because of its specific molecular structure. The spontaneous polarization vector is perpendicular to the c-director and this polarization is produced from the chiral nature of the elongated molecules in smectic layers. The interlayer molecular arrangements rotate with a fixed tilt angle by the application of a sufficiently large electric field and it brings the different layer polarization vectors in a particular orientationwith each other aswell aswith the applied electric field. The transition from AFLCs to ferroelectric liquid crystals (FLCs) occurs only under certain conditions [11].The antiferroelectric configuration loses its stability at a particular field called critical field. At a very low value of the electric field a transition exists other than from antiferroelectric to ferroelectric transition called Freedericksz transition for finite dimensional liquid crystal molecules. Such transition is important [12] because of the creation of dielectric anisotropy of the medium as well as the surface effect. The real part of the relative complex dielectric permittivity depends on the change of polarization of the different adjacent layers with the applied electric field at a particular frequency. The imaginary part of it gives the dielectric losses of the medium for a particular mode of relaxation. Many researchers [13-25] observed different modes of relaxation for a long period and long range of frequencies. Most of the researchers suggest two modes of relaxation, i.e. low frequency mode called in-phase mode and high frequency mode named as anti-phase mode. In earlier publication [26] discussed were both phases in the presence of interlayer interaction in AFLCs and justified was results with the experimental reported above values. The ionic influence [27] was also studied and observed was the variation of elastic properties with the change in ion density. Studied was the dielectric behavior of AFLCs in the presence of free volumes and the surface anchoring strength of polymer matrix for AFLCs [28]. Theoretically authors studied the mechanical contribution [29] that arose due to the flexoelectric effect which is an important part of AFLCs for the construction of devices and theoretically was observed the importance of spin–spin interaction for fabrication purpose [30]. The detailed numerical investigation [31] was done about the importance of the Freedericksz transition in device applications for nematic liquid crystals, the author is reporting an elaborate theoretical model to discuss about the behavior of dielectric functions with the consideration of dielectric anisotropy and surface anchoring effect using the Landau–Ginzburg model in the AFLC system. As noted by the author the dielectric anisotropy is a normal phenomenon of the finite dimensional AFLCs in the presence of the electric field. In the model which is discussed Freedericksz transition of AFLCs considering only nearest neighbor interactions between smectic layerswith the inclusion of dielectric anisotropy theoretically was obtained the variation of the relative dielectric constant, dielectric losses and dielectric strength during Freedericksz transition. Except those both in-phase and anti-phase motions were significantly varied by considering the importance of dielectric anisotropy using the Landau–Ginzburg equation for a purely AFLC system. Recently [32] antiferroelectric liquid crystal/carbon nano tube duo was studied for achieving modified electro-optical properties aiming at display applications. The author finds that multi-walled carbon nanotubes (MWCNT) are dispersed in a high tilted antiferroelectric liquid crystal composed of rod like molecules. The effects of nano-dispersion on electro-optical and dielectric properties of the host were studied in details. The time for switching between dark and bright states and the rotational viscosity are reduced and spontaneous polarization is enhanced considerably by minute addition of MWCNTs. The high tilt angle of the molecules necessary for obtaining good dark state in displays has not changed after dispersion of nanotubes. The strong interaction of the aromatic cores of the rod like liquid crystal molecules with the honey-comb pattern of the CNT walls is considered responsible behind such improvements of physical properties of the host. Recently nano-dispersion studies have been done using ferroelectric liquid crystals (FLC) as the host medium [33-39].The response time is found to reduce considerably by addition of multi-walled carbon nanotubes (MWCNTs) in the FLC medium and the fastness of response is attributed to the decrease in rotational viscosity and increase in anchoring strength [20]. As noted in [32] the study of multi walled carbon nanotubes dispersed in antiferroelectric liquid crystal leads to composed of rod like molecules (AFLC). It was by these authors successfully demonstrated that minute addition of CNTs can reduce the response time for switching between bright and dark states by decreasing the rotational viscosity of an AFLC system. Tilt angle is essentially unaltered which is necessary to achieve perfect dark state of an AFLC device. The spontaneous polarization has been increased owing to the strong interaction of the aromatic cores of the rod like liquid crystal molecules with the honey-comb pattern of the CNT walls The *SmC*A* ordering is improved around the CNTs. All relaxation modes in each phase are facilitated by dispersion of nanotubes whicg enhances the relaxation frequencies. This composite system with improvement in electro-optical properties can be exploited in modern display applications. *SmC** LC were studied by one of t he authors (OH) [40 – 41], namely their transition from the *SmC* phase. The role of disclinations in an applied electric field was studied by O. Hudak [42] in the *SmC** structure. While the role of fluctuations and of disclinations in the *SmC** (FLC) was studied in the above papers [40 – 42], the role of fluctuations and disclinations in *SmC\*A* and the role of addition of MWCNTs in *SmC\*A* phaseis not well understood. Our theoretical description of (AFLC) including *SmC\*A* introduced in [1] may thus be used to study these effects.

Bibliography.